\documentclass[aps,prd,preprint,superscriptaddress]{revtex4-1}
\usepackage{latexsym}
\usepackage{amsmath,amssymb}
\usepackage{graphicx}
\usepackage{xcolor}
\usepackage{cancel}  % \cancel{A}, \bcancel{A}, \xcancel{A}, \cancelto{0}{A}

\usepackage{hyperref}     % color hypertext for pdflatex
\hypersetup{colorlinks,%
  citecolor=blue,%
  linkcolor=cyan,%
  pdftex}

\begin{document}

%\preprint{arXiv:yymm.nnnn [hep-th]}

\title{Remarks on generalized uncertainty principle induced from constraint system}

%%%%%%%%%%%%%%%%%%%% PRD Author Style
\author{Myungseok Eune} %
\email[]{younms@sogang.ac.kr} %
\affiliation{Research Institute for Basic Science, Sogang University,
  Seoul, 121-742, Republic of Korea} %

\author{Wontae Kim} %
\email[]{wtkim@sogang.ac.kr} %
\affiliation{Research Institute for Basic Science, Sogang University,
  Seoul, 121-742, Republic of Korea} %
\affiliation{Department of Physics, Sogang University, Seoul 121-742,
  Republic of Korea} %
\affiliation{Center for Quantum Spacetime, Sogang University, Seoul
  121-742, Republic of Korea} %

\date{\today}

\begin{abstract}
  The extended commutation relations for a generalized uncertainty
  principle have been based on the assumption of the minimal length in
  position.  Instead of this assumption, we start with a constrained
  Hamiltonian system described by the conventional Poisson algebra and
  then impose appropriate second class constraints to this system.
  Consequently, we can show that the consistent Dirac brackets for
  this system are nothing but the extended commutation relations
  describing the generalized uncertainty principle.
\end{abstract}

%\pacs{03.65.-w}
%\keywords{Emergence of cosmic space}

\maketitle

%%%%%%%%%%%%%%%%%%%%%%%%%%%%%%%%%%%%%%%%%%%%%%%%

%% 1 Introduction to GUP and its applications
The realization of quantum gravity is one of the major subjects in
high energy physics, however, there has been difficulties to obtain a
renormalizable Einstein's gravity in spite of intensive studies to
construct a renormalizable theory of gravity.  All divergences from
quantum corrections can be controlled by a cutoff so that a minimal
length has been naturally introduced in string
theory~\cite{Maggiore:1993kv,Maggiore:1993zu,Garay:1994en}.  In the
context of quantum mechanics, this minimal length can be obtained
systemically from a generalized uncertainty principle
(GUP)~\cite{Kempf:1993bq,Kempf:1994su,Kempf:1996fz,Chang:2001kn,Chang:2001bm}
which is a modification of the Heisenberg's uncertainty principle.  In
one dimensional space, the simplest GUP with a minimal length is given
by \(\Delta x \Delta p \ge (\hbar/2) \left[1 + \beta (\Delta p)^2
\right]\), where $\Delta x$ and $\Delta p$ are uncertainties for
positon and momentum, respectively, and $\beta$ is a positive constant
which is independent of $\Delta x$ and $\Delta p$. It can be easily
shown that GUP has the minimal length \(\Delta x_{\rm min} =
\hbar\sqrt{\beta}\).  A commutation relation leading to GUP can be
written as \([x, p]_{\rm GUP} = i\hbar(1+\beta p^2)\), where $x$ and
$p$ are the positon and the momentum operators, respectively.

On the other hand, any Hamiltonian system on the GUP can be treated as
the perturbed canonical Hamiltonian system~\cite{Das:2008kaa,
  Ali:2009zq, Das:2009hs, Basilakos:2010vs, Chemissany:2011nq,
  Ali:2011fa}. This GUP-corrected term of the Hamiltonian gives
quantum gravity corrections to various quantum phenomena.  It has been
also studied that the noncommutative algebra can be generated from the
Poisson brackets of the Lorentz-covariant observables in some
models~\cite{Deriglazov:2012xc}.  Recently, the covariant form of the
GUP algebra proposed in Ref.~\cite{Kempf:1996nk}, which is different
from the above GUP algebra, has been interpreted as the Dirac bracket
derived from a constrained symplectic
structure~\cite{Pramanik:2013zy}.

On general grounds, in $D$-dimensional space the one-dimensional
commutation relation can be extended to the tensorial form \([x_i,
p_j]_{\rm GUP} = i\hbar [(1+\beta p^2) \delta_{ij} + \beta' p_i
p_j]\)~\cite{Kempf:1994su,Kempf:1996fz,Chang:2001bm}, where \(i,j = 1,
\cdots, D\).  If the momentum operator is assumed to commute with each
other, \([p_i, p_j]_{\rm GUP} = 0\), then the Jacobi identity
determines the commutation relations among the coordinates as \([x_i,
x_j]_{\rm GUP} = i\hbar [(2\beta - \beta' + (2\beta + \beta') \beta
p^2)/(1+\beta p^2)] (p_i x_j - p_j x_i)\).  These commutation
relations can be rewritten in terms of Poisson brackets for
simplicity,
\begin{align}
  \{x_i, x_j \}_{\rm GUP} &= \frac{2\beta - \beta' + (2\beta + \beta')
    \beta
    p^2}{1+\beta p^2} (p_i x_j - p_j x_i). \label{gup:xx} \\
  \{x_i, p_j\}_{\rm GUP} &= (1+\beta p^2) \delta_{ij} + \beta' p_i
  p_j, \label{gup:xp} \\
  \{p_i, p_j\}_{\rm GUP} &= 0, \label{gup:pp}
\end{align}
which can be reduced to the conventional Poisson brackets when
$\beta=\beta' =0$. It means that the conventional Poisson brackets can
be deformed by something, which eventually yields the GUP brackets.
In other words, one can regard the GUP brackets as the Dirac brackets
\cite{Dirac:1950pj} which are responsible for the second class
constraint system. 

In this paper, we would like to derive the above GUP brackets
(commutators) in terms of the Dirac method, which is the motivation of
the present work.  For this purpose, we will consider additional
degrees of freedom of $\tilde{x}_i$ and $\tilde{p}_i$ where the total
number of degrees of freedom is tentatively doubled in phase space,
and then impose special second class constraints $\Omega^{(1)}_i =
\Omega^{(1)}_i (x_i,p_i, \tilde{x_i}, \tilde{p_i}) \approx 0$ and
$\Omega^{(2)}_i= \Omega^{(2)}_i (x_i,p_i, \tilde{x_i} ,
\tilde{p_i})\approx 0$ in order to remove the additional degrees of
freedom in a nontrivial way. Eventually, it remarkably gives the
expression~\eqref{gup:xx}--\eqref{gup:pp} of GUP brackets in the
reduced phase space with the original variables.  Consequently, we
shall find that the number of degrees of freedom is unchanged and the
equations of motion are the same with those of GUP.

Let us assume a canonical Hamiltonian system which is described by
\begin{equation}
 H_c = \frac{p^2}{2m} + V(x), \label{c}
\end{equation}
without loss of generality.  Then, we consider an extended phase space
which consists of old variables $(x_i, p_i)$ and new variables
$(\tilde{x_i}, \tilde{p_i})$.  In this enlarged phase space, the
variables should be constrained, and one can specify some special
surfaces to yield the nontrivial brackets. It is based on the standard
lore of Dirac method which tells us that the conventional Poisson
brackets can be modified by the second class constraints. We hope that
the modified brackets turns out to be GUP brackets. Now, the essential
point is how to figure out the second class constraints. By trial and
error, we make an ansatz for the second class constraints which can be
written in the form of
\begin{align}
  \Omega^{(1)}_i &= \tilde{x}_i + \beta p^2 (x_i + \tilde{x}_i) +
  \beta' p_i \, p\cdot(x + \tilde{x}) \approx 0, \label{Omega:1}\\
  \Omega^{(2)}_i &= p_i - \tilde{p}_i \approx 0. \label{Omega:2}
\end{align}
Then, the primary Hamiltonian implemented by the second class
constraints is given by
\begin{align}
  H_p &= \frac{p^2}{2m} + V(x) + \sum_{a=1}^2 u^{(a)}_i \Omega^{(a)}_i
  (x,p, \tilde{x}, \tilde{p}), \label{H:p}
\end{align}
where $V(x)$ and $u^{(a)}_i$ are the potential energy and the
multipliers. Note that the additional variables does not appears in
the part of the kinetic energy since it should be nondynamical so that
they are embedded in the potential.  The constraints~\eqref{Omega:1}
and~\eqref{Omega:2} form a second class constraint algebra since their
Poisson brackets \(C_{ij}^{(ab)} \equiv \{\Omega_i^{(a)},
\Omega_j^{(b)} \}\) are given by \(C_{ij}^{(11)} = (2\beta + \beta')
\beta p^2 \gamma_{ij}\), \(C_{ij}^{(12)} = - C_{ij}^{(21)} = -
\delta_{ij}\), and \(C_{ij}^{(22)} = 0\), where \(\gamma_{ij} \equiv
p_i(x_j + \tilde{x}_j) - p_j (x_i + \tilde{x}_i) \). Using the
constraint~\eqref{Omega:1}, one can obtain the relation \( \gamma_{ij}
=[1/(1+\beta p^2)] (p_i x_j - p_j x_i) \).  The Dirac bracket is
defined by \(\{A, B\}_{\rm D} = \{A, B\} - \sum_{a,b} \{A,
\Omega_i^{(a)} \} (C^{-1})_{ij}^{(ab)} \{\Omega_j^{(b)}, B\} \), where
\(\{, \}\) is the conventional Poisson bracket. Then, the Dirac
brackets between $x_i$ and $p_i$ are obtained as
\begin{align}
   \{x_i, x_j \}_{\rm D} &= \frac{2\beta - \beta' + (2\beta + \beta')
    \beta
    p^2}{1+\beta p^2} (p_i x_j - p_j x_i), \label{Dirac:xx} \\ 
  \{x_i, p_j \}_{\rm D} &= (1+\beta p^2) \delta_{ij} + \beta'
  p_i p_j, \label{Dirac:xp}\\
  \{p_i, p_j \}_{\rm D} &= 0. \label{Dirac:pp}
\end{align}
%and the brackets related to the auxiliary coordinates and momenta are
%given by
%\begin{align}
%  \{\tilde{x}_i, \tilde{x}_j \}_{\rm D} &= (2\beta +
%  \beta') \beta p^2 \gamma_{ij}, \label{Dirac:x2x2}\\
%  \{\tilde{x}_i, \tilde{p}_j \}_{\rm D} &= -\beta p^2 \delta_{ij} -
%  \beta' p_i p_j, \label{Dirac:x2p2} \\
%  \{\tilde{p}_i, \tilde{p}_j \}_{\rm D} &= 0, \label{Dirac:p2p2} \\
%  \{x_i, \tilde{x}_j \}_{\rm D} &= - 2\beta p_i (x_j + \tilde{x}_j) -
%  (2\beta + \beta') \beta p^2 \gamma_{ij} - \beta'[p_j(x_i +
%  \tilde{x}_i) + p \cdot (x+\tilde{x})
%  \delta_{ij}], \label{Dirac:x.x2}\\
%  \{x_i, \tilde{p}_j \}_{\rm D} &= \{x_i, p_j \}_{\rm D} =
%  (1+\beta p^2) \delta_{ij} + \beta' p_i p_j, \label{Dirac:x.p2} \\
%  \{\tilde{x}_i, p_j \}_{\rm D} &= \{\tilde{x}_i, \tilde{p}_j \}_{\rm
%    D} = -\beta p^2 \delta_{ij} - \beta' p_i p_j, \label{Dirac:x2.p} \\
%  \{p_i, \tilde{p}_j \}_{\rm D} &= 0. \label{Dirac:p.p2}
%\end{align}
Thus, it can be shown that Eqs.~\eqref{Dirac:xx}, \eqref{Dirac:xp},
and~\eqref{Dirac:pp} are equivalent to Eqs.~\eqref{gup:xx},
\eqref{gup:xp}, and~\eqref{gup:pp}, respectively. The reduced
Hamiltonian can be obtained by strongly imposing the
constraints~\eqref{Omega:1} and \eqref{Omega:2} which eliminate
$\tilde{x}_i$ and $\tilde{p}_i$ in the Hamiltonian, which recover the
canonical Hamiltonian \eqref{c}.
%%%%%%%%%%%%%%%%%%%%%%%%%%%%%%%%%%%%%%%%%%%%%%%%%
So, the Hamilton's equations are obtained as
\begin{align}
  \dot{x}_i &= \{x_i, H_c \}_{\rm D} \notag \\
  &=[1+(\beta + \beta') p^2] \frac{p_i}{m} + \frac{2\beta -
    \beta' + (2\beta + \beta') \beta p^2}{1+\beta p^2} (p_i x_j - p_j
  x_i) \frac{\partial V}{\partial x_j}, \label{H:eq:dot.x}\\
  \dot{p}_i &= \{p_i, H_c \}_{\rm D} \notag\\
  &= - [(1+\beta p^2) \delta_{ij} + \beta' p_i p_j] \frac{\partial
    V}{\partial x_j},% = \dot{\tilde{p}}_i, 
	\label{H:eq:dot.p} %\\
%  \dot{\tilde{x}}_i &= \{\tilde{x}_i, H_p \}_{\rm D} \notag \\
%  &= [2\beta p_j(x_i + \tilde{x}_i) + \beta' (p_i (x_j + \tilde{x}_j)
%  + p\cdot(x+\tilde{x}) \delta_{ij}) -(2\beta + \beta') \beta p^2
%  \gamma_{ij}] \frac{\partial V}{\partial x_j} - (\beta + \beta')
%  \frac{p^2 p_i}{m}, \label{H:eq:dot.x2}
\end{align}
where the overdot denotes the derivative with respect to time. 

In conclusion, we have obtained the GUP brackets (commutators) which
are related to the minimal length from the second class constraint
system, and identified the GUP brackets with the Dirac brackets. It
means that the Hamiltonian system described by the GUP brackets can be
interpreted as a gauge fixed version of the first class constraint
system. The similar behaviors can be found in the chiral Schwinger
model~\cite{Harada:1986tc,Kim:1992ey,Natividade:2000sh} and the
Chern-Simons theory~\cite{Banerjee:1993pm,Kim:1994np,Banerjee:1996sp}.
% Formally it is possible to make the second
% class constraint system to be the first class constraint system by
% adding additional degrees of freedom systematically in terms of
% Batalin-Tyutin formalism~\cite{Batalin:1986aq, Batalin:1986fm,
%   Batalin:1991jm}; however it is hard to realize it in this model
% because of complexity of constraint structure unfortunately.

 %%%%%% Acknowledgments %%%%%%%%%
\begin{acknowledgments}
  We would like to thank Edwin J.\ Son for exciting discussions.  This
  work was supported by the Sogang University Research Grant of
  201310022 (2013).
\end{acknowledgments}

%%%%%%%%%%%%%%%%%%%%%%%%%%%%%%%%%%%%%%%%%%%%%%%%
%%%%%%%%%%%%%%%             References         %%%%%%%%%%%%%%%%
%%%%%%%%%%%%%%%%%%%%%%%%%%%%%%%%%%%%%%%%%%%%%%%%

% Create the reference section using BibTeX:
%\bibliography{basename of .bib file} 
%\bibliographystyle{mybib}
\bibliographystyle{apsrev} % PRD
\bibliography{references}

\begin{thebibliography}{24}
\expandafter\ifx\csname natexlab\endcsname\relax\def\natexlab#1{#1}\fi
\expandafter\ifx\csname bibnamefont\endcsname\relax
  \def\bibnamefont#1{#1}\fi
\expandafter\ifx\csname bibfnamefont\endcsname\relax
  \def\bibfnamefont#1{#1}\fi
\expandafter\ifx\csname citenamefont\endcsname\relax
  \def\citenamefont#1{#1}\fi
\expandafter\ifx\csname url\endcsname\relax
  \def\url#1{\texttt{#1}}\fi
\expandafter\ifx\csname urlprefix\endcsname\relax\def\urlprefix{URL }\fi
\providecommand{\bibinfo}[2]{#2}
\providecommand{\eprint}[2][]{\url{#2}}

\bibitem[{\citenamefont{Maggiore}(1993)}]{Maggiore:1993kv}
\bibinfo{author}{\bibfnamefont{M.}~\bibnamefont{Maggiore}},
  \bibinfo{journal}{Phys.\ Lett.\ B} \textbf{\bibinfo{volume}{319}},
  \bibinfo{pages}{83} (\bibinfo{year}{1993}), \eprint{hep-th/9309034}.

\bibitem[{\citenamefont{Maggiore}(1994)}]{Maggiore:1993zu}
\bibinfo{author}{\bibfnamefont{M.}~\bibnamefont{Maggiore}},
  \bibinfo{journal}{Phys.\ Rev.\ D} \textbf{\bibinfo{volume}{49}},
  \bibinfo{pages}{5182} (\bibinfo{year}{1994}), \eprint{hep-th/9305163}.

\bibitem[{\citenamefont{Garay}(1995)}]{Garay:1994en}
\bibinfo{author}{\bibfnamefont{L.~J.} \bibnamefont{Garay}},
  \bibinfo{journal}{Int.\ J.\ Mod.\ Phys.\ A} \textbf{\bibinfo{volume}{10}},
  \bibinfo{pages}{145} (\bibinfo{year}{1995}), \eprint{gr-qc/9403008}.

\bibitem[{\citenamefont{Kempf}(1994)}]{Kempf:1993bq}
\bibinfo{author}{\bibfnamefont{A.}~\bibnamefont{Kempf}}, \bibinfo{journal}{J.\
  Math.\ Phys.} \textbf{\bibinfo{volume}{35}}, \bibinfo{pages}{4483}
  (\bibinfo{year}{1994}), \eprint{hep-th/9311147}.

\bibitem[{\citenamefont{Kempf et~al.}(1995)\citenamefont{Kempf, Mangano, and
  Mann}}]{Kempf:1994su}
\bibinfo{author}{\bibfnamefont{A.}~\bibnamefont{Kempf}},
  \bibinfo{author}{\bibfnamefont{G.}~\bibnamefont{Mangano}}, \bibnamefont{and}
  \bibinfo{author}{\bibfnamefont{R.~B.} \bibnamefont{Mann}},
  \bibinfo{journal}{Phys.\ Rev.\ D} \textbf{\bibinfo{volume}{52}},
  \bibinfo{pages}{1108} (\bibinfo{year}{1995}), \eprint{hep-th/9412167}.

\bibitem[{\citenamefont{Kempf}(1997)}]{Kempf:1996fz}
\bibinfo{author}{\bibfnamefont{A.}~\bibnamefont{Kempf}}, \bibinfo{journal}{J.\
  Phys.\ A} \textbf{\bibinfo{volume}{30}}, \bibinfo{pages}{2093}
  (\bibinfo{year}{1997}), \eprint{hep-th/9604045}.

\bibitem[{\citenamefont{Chang et~al.}(2002{\natexlab{a}})\citenamefont{Chang,
  Minic, Okamura, and Takeuchi}}]{Chang:2001kn}
\bibinfo{author}{\bibfnamefont{L.~N.} \bibnamefont{Chang}},
  \bibinfo{author}{\bibfnamefont{D.}~\bibnamefont{Minic}},
  \bibinfo{author}{\bibfnamefont{N.}~\bibnamefont{Okamura}}, \bibnamefont{and}
  \bibinfo{author}{\bibfnamefont{T.}~\bibnamefont{Takeuchi}},
  \bibinfo{journal}{Phys.\ Rev.\ D} \textbf{\bibinfo{volume}{65}},
  \bibinfo{pages}{125027} (\bibinfo{year}{2002}{\natexlab{a}}),
  \eprint{hep-th/0111181}.

\bibitem[{\citenamefont{Chang et~al.}(2002{\natexlab{b}})\citenamefont{Chang,
  Minic, Okamura, and Takeuchi}}]{Chang:2001bm}
\bibinfo{author}{\bibfnamefont{L.~N.} \bibnamefont{Chang}},
  \bibinfo{author}{\bibfnamefont{D.}~\bibnamefont{Minic}},
  \bibinfo{author}{\bibfnamefont{N.}~\bibnamefont{Okamura}}, \bibnamefont{and}
  \bibinfo{author}{\bibfnamefont{T.}~\bibnamefont{Takeuchi}},
  \bibinfo{journal}{Phys.\ Rev.\ D} \textbf{\bibinfo{volume}{65}},
  \bibinfo{pages}{125028} (\bibinfo{year}{2002}{\natexlab{b}}),
  \eprint{hep-th/0201017}.

\bibitem[{\citenamefont{Das and Vagenas}(2008)}]{Das:2008kaa}
\bibinfo{author}{\bibfnamefont{S.}~\bibnamefont{Das}} \bibnamefont{and}
  \bibinfo{author}{\bibfnamefont{E.~C.} \bibnamefont{Vagenas}},
  \bibinfo{journal}{Phys.\ Rev.\ Lett.} \textbf{\bibinfo{volume}{101}},
  \bibinfo{pages}{221301} (\bibinfo{year}{2008}), \eprint{0810.5333}.

\bibitem[{\citenamefont{Ali et~al.}(2009)\citenamefont{Ali, Das, and
  Vagenas}}]{Ali:2009zq}
\bibinfo{author}{\bibfnamefont{A.~F.} \bibnamefont{Ali}},
  \bibinfo{author}{\bibfnamefont{S.}~\bibnamefont{Das}}, \bibnamefont{and}
  \bibinfo{author}{\bibfnamefont{E.~C.} \bibnamefont{Vagenas}},
  \bibinfo{journal}{Phys.\ Lett.\ B} \textbf{\bibinfo{volume}{678}},
  \bibinfo{pages}{497} (\bibinfo{year}{2009}), \eprint{0906.5396}.

\bibitem[{\citenamefont{Das and Vagenas}(2009)}]{Das:2009hs}
\bibinfo{author}{\bibfnamefont{S.}~\bibnamefont{Das}} \bibnamefont{and}
  \bibinfo{author}{\bibfnamefont{E.~C.} \bibnamefont{Vagenas}},
  \bibinfo{journal}{Canad.\ J.\ Phys.} \textbf{\bibinfo{volume}{87}},
  \bibinfo{pages}{233} (\bibinfo{year}{2009}), \eprint{0901.1768}.

\bibitem[{\citenamefont{Basilakos et~al.}(2010)\citenamefont{Basilakos, Das,
  and Vagenas}}]{Basilakos:2010vs}
\bibinfo{author}{\bibfnamefont{S.}~\bibnamefont{Basilakos}},
  \bibinfo{author}{\bibfnamefont{S.}~\bibnamefont{Das}}, \bibnamefont{and}
  \bibinfo{author}{\bibfnamefont{E.~C.} \bibnamefont{Vagenas}},
  \bibinfo{journal}{JCAP} \textbf{\bibinfo{volume}{1009}}, \bibinfo{pages}{027}
  (\bibinfo{year}{2010}), \eprint{1009.0365}.

\bibitem[{\citenamefont{Chemissany et~al.}(2011)\citenamefont{Chemissany, Das,
  Ali, and Vagenas}}]{Chemissany:2011nq}
\bibinfo{author}{\bibfnamefont{W.}~\bibnamefont{Chemissany}},
  \bibinfo{author}{\bibfnamefont{S.}~\bibnamefont{Das}},
  \bibinfo{author}{\bibfnamefont{A.~F.} \bibnamefont{Ali}}, \bibnamefont{and}
  \bibinfo{author}{\bibfnamefont{E.~C.} \bibnamefont{Vagenas}},
  \bibinfo{journal}{JCAP} \textbf{\bibinfo{volume}{1112}}, \bibinfo{pages}{017}
  (\bibinfo{year}{2011}), \eprint{1111.7288}.

\bibitem[{\citenamefont{Ali et~al.}(2011)\citenamefont{Ali, Das, and
  Vagenas}}]{Ali:2011fa}
\bibinfo{author}{\bibfnamefont{A.~F.} \bibnamefont{Ali}},
  \bibinfo{author}{\bibfnamefont{S.}~\bibnamefont{Das}}, \bibnamefont{and}
  \bibinfo{author}{\bibfnamefont{E.~C.} \bibnamefont{Vagenas}},
  \bibinfo{journal}{Phys.\ Rev.\ D} \textbf{\bibinfo{volume}{84}},
  \bibinfo{pages}{044013} (\bibinfo{year}{2011}), \eprint{1107.3164}.

\bibitem[{\citenamefont{Deriglazov}(2012)}]{Deriglazov:2012xc}
\bibinfo{author}{\bibfnamefont{A.}~\bibnamefont{Deriglazov}},
  \bibinfo{journal}{Phys.\ Lett.\ A} \textbf{\bibinfo{volume}{377}},
  \bibinfo{pages}{13} (\bibinfo{year}{2012}), \eprint{1203.5697}.

\bibitem[{\citenamefont{Kempf and Mangano}(1997)}]{Kempf:1996nk}
\bibinfo{author}{\bibfnamefont{A.}~\bibnamefont{Kempf}} \bibnamefont{and}
  \bibinfo{author}{\bibfnamefont{G.}~\bibnamefont{Mangano}},
  \bibinfo{journal}{Phys.\ Rev.\ D} \textbf{\bibinfo{volume}{55}},
  \bibinfo{pages}{7909} (\bibinfo{year}{1997}), \eprint{hep-th/9612084}.

\bibitem[{\citenamefont{Pramanik and Ghosh}(2013)}]{Pramanik:2013zy}
\bibinfo{author}{\bibfnamefont{S.}~\bibnamefont{Pramanik}} \bibnamefont{and}
  \bibinfo{author}{\bibfnamefont{S.}~\bibnamefont{Ghosh}}
  (\bibinfo{year}{2013}), \eprint{1301.4042}.

\bibitem[{\citenamefont{Dirac}(1950)}]{Dirac:1950pj}
\bibinfo{author}{\bibfnamefont{P.~A.} \bibnamefont{Dirac}},
  \bibinfo{journal}{Canad.\ J.\ Math.} \textbf{\bibinfo{volume}{2}},
  \bibinfo{pages}{129} (\bibinfo{year}{1950}).

\bibitem[{\citenamefont{Harada and Tsutsui}(1987)}]{Harada:1986tc}
\bibinfo{author}{\bibfnamefont{K.}~\bibnamefont{Harada}} \bibnamefont{and}
  \bibinfo{author}{\bibfnamefont{I.}~\bibnamefont{Tsutsui}},
  \bibinfo{journal}{Prog.Theor.Phys.} \textbf{\bibinfo{volume}{78}},
  \bibinfo{pages}{878} (\bibinfo{year}{1987}).

\bibitem[{\citenamefont{Kim et~al.}(1992)\citenamefont{Kim, Kim, Kim, Park,
  Kim, and Kim}}]{Kim:1992ey}
\bibinfo{author}{\bibfnamefont{Y.-W.} \bibnamefont{Kim}},
  \bibinfo{author}{\bibfnamefont{S.-K.} \bibnamefont{Kim}},
  \bibinfo{author}{\bibfnamefont{W.-T.} \bibnamefont{Kim}},
  \bibinfo{author}{\bibfnamefont{Y.-J.} \bibnamefont{Park}},
  \bibinfo{author}{\bibfnamefont{K.~Y.} \bibnamefont{Kim}}, \bibnamefont{and}
  \bibinfo{author}{\bibfnamefont{Y.}~\bibnamefont{Kim}},
  \bibinfo{journal}{Phys.\ Rev.\ D} \textbf{\bibinfo{volume}{46}},
  \bibinfo{pages}{4574} (\bibinfo{year}{1992}).

\bibitem[{\citenamefont{Natividade et~al.}(2004)\citenamefont{Natividade,
  Boschi-Filho, and Belvedere}}]{Natividade:2000sh}
\bibinfo{author}{\bibfnamefont{C.}~\bibnamefont{Natividade}},
  \bibinfo{author}{\bibfnamefont{H.}~\bibnamefont{Boschi-Filho}},
  \bibnamefont{and}
  \bibinfo{author}{\bibfnamefont{L.}~\bibnamefont{Belvedere}},
  \bibinfo{journal}{Mod.\ Phys.\ Lett.\ A} \textbf{\bibinfo{volume}{19}},
  \bibinfo{pages}{2957} (\bibinfo{year}{2004}), \eprint{hep-th/0010081}.

\bibitem[{\citenamefont{Banerjee}(1993)}]{Banerjee:1993pm}
\bibinfo{author}{\bibfnamefont{R.}~\bibnamefont{Banerjee}},
  \bibinfo{journal}{Phys.\ Rev.\ D} \textbf{\bibinfo{volume}{48}},
  \bibinfo{pages}{5467} (\bibinfo{year}{1993}).

\bibitem[{\citenamefont{Kim and Park}(1994)}]{Kim:1994np}
\bibinfo{author}{\bibfnamefont{W.~T.} \bibnamefont{Kim}} \bibnamefont{and}
  \bibinfo{author}{\bibfnamefont{Y.-J.} \bibnamefont{Park}},
  \bibinfo{journal}{Phys.\ Lett.\ B} \textbf{\bibinfo{volume}{336}},
  \bibinfo{pages}{376} (\bibinfo{year}{1994}), \eprint{hep-th/9403188}.

\bibitem[{\citenamefont{Banerjee et~al.}(1997)\citenamefont{Banerjee, Rothe,
  and Rothe}}]{Banerjee:1996sp}
\bibinfo{author}{\bibfnamefont{R.}~\bibnamefont{Banerjee}},
  \bibinfo{author}{\bibfnamefont{H.}~\bibnamefont{Rothe}}, \bibnamefont{and}
  \bibinfo{author}{\bibfnamefont{K.}~\bibnamefont{Rothe}},
  \bibinfo{journal}{Phys.\ Rev.\ D} \textbf{\bibinfo{volume}{55}},
  \bibinfo{pages}{6339} (\bibinfo{year}{1997}), \eprint{hep-th/9611077}.

\end{thebibliography}

%%%%%%%%%%%%%%%%%%%%%%%%%%%%%%%%%%%%%%%%%%%%%%%%%%%
%%%%%%%%%%%%%%%             References (.bib)      %%%%%%%%%%%%%%%%
%%%%%%%%%%%%%%%%%%%%%%%%%%%%%%%%%%%%%%%%%%%%%%%%%%%

%\begin{filecontents}{references.bib}
%
%
%\end{filecontents}

\end{document}